\documentstyle[12pt]{article}
\begin{document}

\setlength{\baselineskip}{0.30in}
\def\psl{p \hspace*{-0.5em}/}
\def\ksl{k \hspace*{-0.5em}/}
\newcommand{\be}{\begin{equation}}
\newcommand{\bi}{\bibitem}
\newcommand{\ee}{\end{equation}}
\newcommand{\ra}{\rightarrow}

\begin{flushright}
UM-TH-00-08 \\ 
Freiburg-THEP-007\\
\end{flushright}

\begin{center}
\vglue .06in
{\Large \bf {A Note on Large Extra Dimensions.}}\\ [.5in]

{\bf R. Akhoury$^{(1)}$ and J.J. van der Bij$^{(1,2)}$}\\ [.15in]

{\it $^{(1)}$ The Randall Laboratory of Physics\\
University of Michigan\\
Ann Arbor, MI 48109-1120}\\ [.15in]
{\it and}\\ [.15in]
{\it $^{(2)}$ Fakult\"at f\"ur Physik\\
Universit\"at Freiburg\\
H. Herderstr. 3, 79104 Freiburg\\}
\end{center}
\begin{abstract}
\begin{quotation}
We study corrections to the photon-propagator in a recently
proposed model, where gravity lives in 4+n dimensions and Standard Model fields
in 4 dimensions. We find a correction to the formfactor of the
photon that can be constrained by QED tests.
\end{quotation}
\end{abstract}
\newpage

\section{Introduction }
Motivated in part by naturalness issues, recently, starting with ref. \cite{anton}, there have
been several  attempts to develop models where the effects of gravity  become large at energy
scales of the order of a TeV~-~ much lower than the traditionally accepted Planck
mass.  In fact, in a previous paper\cite{bij1} by one of the authors it was
suggested that the so-called hierarchy problem, the question why the Planck mass is
much larger than the weak scale, could be solved by having gravity play a role at
the electroweak scale. One of the implications of such  a principle is the
possibility of having large extra dimensions being  present. This possibility has
recently been presented in a simple form  in ref.\cite{dim1}. In this paper it was
suggested that the standard  model is confined on a D3 brane, but that gravity lives
in  the bulk of the  (4+n) dimensional space-time,
where the n dimensions are compactified, possibly with a large 
radius. From the formula $M^2_{Pl,4} \approx R^n M^{n+2}_{Pl,4+n}$, the
$4+n$~-~dimensional Planck mass can be at the TeV scale if the size R of the higher 
dimensions is large enough.  As prefigured in ref.\cite{bij1} such
scenarios lead to collider signals in  the form of missing energy
signals and anomalous couplings from higher dimensional operators.
These effects could come from the production of higher dimensional gravitons
and/or their tree-level exchange. Since such signals are absent in
present-day colliders, limits can be put on the scale where the 
effects of higher-dimensional gravity become strong\cite{pheno,peskin}.
Typically one finds limits of the order of a TeV for this scale. Astrophysical
considerations allow one to limit the cases n=1 and 2.\cite{astro}.
 
At first sight the model therefore appears to be in order, at least
for $n>2$. However, since it is non-renormalizable, one has to be
careful that radiative effects do not destroy its consistency.
Non-renormalizable models can often be used as effective Lagrangians,
if the radiative corrections are sufficiently well behaved, so that
the divergent higher-order effects can be absorbed in effective 
local operators, parametrizing our ignorance about the underlying
fundamental dynamics. For example, this procedure works very well
in pion physics, where the starting Lagrangian is the non-linear
sigma-model\cite{weinberg,leutwyler}. However the divergences
in higher-dimensional gravity are much more severe than in the 
non-linear sigma-model, so that explicit calculations are necessary
to determine how far this model makes sense at the quantum level.
Calculations involving loop-effects of higher-dimensional gravitons have
been performed in \cite{taohan,das} for the LEP electroweak precision
data, and in \cite{graesser} for the g-2 factor of the muon.
The authors in \cite{taohan} and \cite{das} both calculate  corrections
to the so called oblique parameters, coming from corrections 
to the vector-boson propagators. They reach different conclusions
about the importance of the corrections, ref.\cite{das} giving
strong constraints, while ref.\cite{taohan} gives much weaker
ones. However both calculations ignore the non-oblique
corrections, which is basically incorrect, since gravity couples to all
particles, not only to vector-bosons. We therefore consider these results
to be too uncertain to provide definite conclusions. In ref.\cite{graesser}
the g-2 factor of the muon was considered. Here it was found that the
corrections are small, well within experimental bounds. However
it is known from studies\cite{bij2} on anomalous vector-boson couplings, that
the g-2 factor of the muon is relatively insensitive to anomalous
effects, that depend on the precise ultraviolet behaviour. The 
form-factor of the photon is a more sensitive quantity, therefore
in the next section we calculate the radiative corrections to 
the photon-propagator due to higher-dimensional graviton exchange.
 
\section{The Calculation}
Such a calculation can be performed using the formalism of ref.\cite{lyk}.
The 4+n dimensional graviton-field Lagrangian is linearized
and expanded in normal modes. The normal modes describe an infinity
of 4-dimensional massive spin-2, spin-1 and spin-0 fields. The spin-1 fields
decouple from ordinary matter, the spin-2 fields couple to the energy-momentum
tensor $T_{\mu \nu}$ and the scalar fields to the trace of the energy
momentum tensor. As we are interested in the coupling to photons, we can
ignore the scalars completely, since the photon energy-momentum tensor
is traceless. The propagator of the massive gravitons is derived in
\cite{veltman}. Because we assume the extra dimensions to be large,
the sum over the modes can be replaced by a density-integral. Assuming
all extra dimensions to be circles with a radius R, the sum over graviton 
modes is replaced by the integral $\int dm^2 R^n m^{n-2} /
((4\pi)^{n/2} \Gamma (n/2))$. 
At the one-loop level there are two types of graphs contributing. The
tadpole graphs contain the 2-photon-2-graviton vertex, which is proportional
to $k_{\mu} k_{\nu} - k^2 \delta_{\mu \nu}$. Such diagrams, contain no further 
momentum dependence. Their contribution can therefore be absorbed in a wave-function
renormalization of the photon. The remaining class of diagrams is non-trivial and
contains the 2-photon-1-graviton vertices. It is strongly ultraviolet divergent and
for low enough dimensions also infra-red divergent. As one also has to take care in
the regularization to preserve the gauge-invariance  of the photon, it is not very
practical to evaluate  this diagram with Feynman parameters. We therefore choose to
evaluate the diagram by a dispersion relation. This way one preserves the symmetries
that have to be present in the theory and at the same time keeps the physically
relevant divergences.  The form of the imaginary part is  determined without
ambiguities. One has however to choose an ultraviolet cut-off $\Lambda$  for the
spectrum, where one assumes new physics takes over. The infrared cut-off $m_{gr}$
is taken as the mass of the lightest graviton.

With these assumptions the imaginary part of the photon two-point
function becomes ($k^2=~-s$):

\begin{eqnarray}
Im \Pi_{\mu \nu}(s) = (\delta_{\mu \nu} k^2 -k_{\mu} k_{\nu})
(- \frac {1}{9 M_{Pl,4+n}^{n+2}}) \times     
( \frac {480 s^{(n+2)/2}}{(n-4)(n+2)(n+4)(n+6)} - \nonumber \\
\frac{s^3}{n-4}m_{gr}^{n-4} 
+\frac{10}{n+2}m_{gr}^{n+2}
-\frac{15}{n+4}{m_{gr}^{n+4} \over s}
+\frac{6}{n+6} s^{-2}m_{gr}^{n+6}
 ) 
\end{eqnarray}
As we assume R to be large the last three terms in the above
expression can be ignored.
This expression is gauge invariant. In the following it will be convenient
to introduce as usual the scalar $\Pi(k^2)$ by
\begin{equation}
\Pi_{\mu \nu}(k)~=~(\delta_{\mu \nu} k^2 -k_{\mu} k_{\nu})\Pi(k^2)
\end{equation}
It is for this object which is free of kinematic singularities that a dispersion 
relation can be written. Indeed, 
the real part of $\Pi(k^2)$ can now be determined by the
relation:
\be
Re \Pi(s) =\frac {1}{\pi} 
\times \int_{m_g^2}^{\Lambda^2} \frac {ds'}{s'-s}Im \Pi(s') 
\ee
The full photon propagator ($\Delta'_{\mu \nu}$), in a general gauge is then given by
\begin{equation}
\Delta'_{\mu \nu} ~=~{\delta_{\mu \nu} \over 
k^2 - k^2 \Pi(k^2)} - {k_{\mu}k_{\nu} \over k^2}
{(1-\alpha)\Pi(k^2) + \alpha \over k^2 - k^2 \Pi(k^2)}
\end{equation}
Note that in the Landau gauge ($\alpha= 1$) the above has a particularly simple form:
\be
\Delta'_{\mu \nu}(k) ~=~{(\delta_{\mu \nu} - k_{\mu}k_{\nu} / k^2) \over 
k^2 - k^2 \Pi(k^2)}
\ee
Since the polarization tensor is transverse, no subtractions are needed for the photon 
mass which stays zero. This is the reason for choosing the scalar function $\Pi(k^2)$
as above with its appropriate kinematic factor.
 However, because of the high
powers of s in imaginary $\Pi(k^2)$ a large number
of subtractions would be needed, each corresponding to a
higher dimensional operator, coming from a taylor series expansion 
of $Re \Pi(k^2)$ in powers of $k^2$. The coefficients of these terms are then
not fixed. As we wish to consider the cut-off to be a physical 
quantity, as in ref.[8-10], we want to use the integral to determine
these coefficients. The lowest order, term ($\Pi(0)$) would correspond to a wave-function
renormalization. So the first non-trivial term is the $k^4$ term in the (inverse) photon
two-point function. It is given by $\Pi'(s=0)$.
To be more precise
the coefficient $\beta$ in the contribution to the polarization tensor
~$\beta k^2(\delta_{\mu \nu}k^2 -k_{\mu} k_{\nu})$~ is given by: 

\begin{eqnarray}
n>4~~~~~\beta =  ({\Lambda^n \over 9\pi M_{Pl,4+n}^{n+2}}) 
 {960 \over n(n-4)(n+2)(n+4)(n+6)}  \nonumber \\
~~~~~n=4~~~~~\beta= ({\Lambda^4 \over 18\pi M_{Pl,4+n}^{n+2}})(\ln({\Lambda^2 \over m_g^2}) -
1/2)  
\nonumber \\ 
~~~~~n<4~~~~~~\beta = ({1 \over 9\pi M_{Pl,4+n}^{n+2}}) 
{m_{gr}^{n-4} \over 2(4-n)}\Lambda^4    
\end{eqnarray}

\section{Discussion.}
The results in the above section clarify some of the questions raised 
in the literature. The question has been raised in ref.[8], whether
radiative contributions can actually grow with the cut-off, or whether
they are always suppressed by $1/ M_{Pl,4+n}^2$. We see for large n a
strong cut-off dependence. It is only when we put $\Lambda = M_{Pl,4+n}$,
that the $1/ M_{Pl,4+n}^2$ behaviour arises. In principle the cut-off
$\Lambda$ is the string scale, where higher spin resonances start appearing.
For the field theory calculation of this paper to be sensible this string scale
should be higher than the Planck scale, as otherwise operators
coming from the higher-spin fields would be more important than the
graviton contribution. In ref. (\cite{peskin}) it has been argued that actually
the string scale is lower than the Planck scale by a factor 1.6-3.
If this is indeed the case, one cannot use the radiative corrections reliably
to give limits on the Planck mass. For comparison with experiment, we
will however take $\Lambda = M_{Pl,4+n}$.

The second question is the meaning of the infrared divergences,
in particular the power divergences in $m_{gr}$. In ref.(\cite{das})
the power divergences were present, however it was argued that they should be absent in the
observables because gravity is IR finite.  In ref.(\cite{taohan}) they were regularized, with the
argument that non-perturbative effects would make them disappear. It is not clear to us what
these effects are. In particular, it is disturbing that the imaginary part of the photon 2 point
function is not finite in the limit $m_{gr} \rightarrow 0$ for $n<4$. We feel that this
singularity is not really an infrared one in the sense that it does not necessarily arise due to
long wavelength excitations. The singular behaviour in Eq. (1), for example, comes from the
$k_{\mu} k_{\nu}k_{\lambda}k_{\rho}/m_{gr}^4$ term in the graviton propagator. If all
invariances in the theory were kept intact then such terms do not contribute to singular
behaviour in physical observables, -for example, if current conservation is preserved as in
massive photon QED then a term like $k_{\mu}k_{\nu}/m_{ph}^2$ in the photon propagator does not
give singularities in observables. In the brane scenario, the brane position breaks the higher
dimensional general coordinate invariance. Thus even though conventional Kaluza Klein theories
or string theories are consistent in terms of preserving invariances, it is not clear if the
brane scenarios in their current form are. We believe that the singular behaviour in Eq. (1) is
a reflection of this. That this depends on the number of extra dimensions could then be simply a
reflection of the fact that the bigger the number of extra dimensions, the harder it is for
higher dimensional gravitons to find the D3 brane. 
Comparing with ref.(\cite{das}), we notice
that in the photon structure the power divergences are similar to those in the previously
calculated S,T,U parameters for $n=3$. It is not so clear to us that terms like the second one
in Eq. (1) can be completely ignored even after a proper understanding of the theoretical issues
involved.

 A limit can be given by comparing the QED-test on the renormalized photon-propagator
$\frac{\delta_{\mu \nu}} {k^2 -k^4/ \Lambda_{QED}^2}$. One has as a limit $\Lambda_{QED} \approx
200 GeV$ from colliders. Putting $\beta = 1/\Lambda_{QED}^2$ gives a limit on
the Planck mass. For $n > 3$ this limit is not competitive with
the limits given for the S,T,U parameters in refs.(\cite{das}, \cite{taohan}) or with
the direct invisible energy search. Thus due to the large denominator in Eq.(6)  we get a limit
at best $M_{Pl,4+n} > 30 GeV$. For the case of $n \leq 3$, somewhat more useful limits can be
set. If we naively keep the terms which are singular as $m_{gr}
\rightarrow 0$ for $n<4$ then one gets large values, greater than a TeV, for the Planck mass.

In conclusion, at the moment, theories with large extra dimensions appear to be consistent at
least for large enough $n$. The problem of singularities in physical observables as $m_{gr}
 \rightarrow 0$ needs to be studied at a theoretical level.

\section{Acknowledgements}
This work was supported by the NATO grant CRG 970113, by a University of Michigan
Rackham grant to promote international partnerships and by the US Department of
Energy. We would like to thank M. Einhorn, E. Yao and O. Yakovlev for discussions.


\end{document}